\documentclass[11pt,a4paper]{article}  
\title{Counting functions for branched covers of elliptic curves \\
and quasi-modular forms}
\author{Hiroyuki Ochiai}
\usepackage{amsmath}

\newcommand{\Z}{\mathbf{Z}}
\newcommand{\Q}{\mathbf{Q}}

\newcommand{\C}{\mathbf{C}}

\newcommand{\Aut}{\mathop{{\mathrm{Aut}}}}
\newcommand{\Res}{\mathop{{\mathrm{Res}}}}
\newcommand{\Hom}{\mathop{{\mathrm{Hom}}}}
\newcommand{\QM}{{\mathrm{QM}}}
\newcommand{\Dq}{{D}}
\newcommand{\vari}{{X}}  
\newcommand{\parti}{{\lambda}} 
\newcommand{\sparti}{{\tilde{\parti}}} 
\newcommand{\ps}[2]{{\tilde{p}_{#1}(#2)}}  
\newcommand{\Young}{{{\cal P}_d}}
\newlength{\theoremindent}
\newtheorem{theorem}{\hspace*{\theoremindent}{\bf Theorem}}
\newtheorem{proposition}[theorem]{\hspace*{\theoremindent}{\bf Proposition}}

\newtheorem{lemma}[theorem]{\hspace*{\theoremindent}{\bf Lemma}}

\newcommand{\qed}{\hspace*{\fill} \fbox{\;}\par\bigskip}

\newtheorem{example}[theorem]{\hspace*{\theoremindent}{\bf Example}}

\date{}
\begin{document}
\maketitle


Abstract:
We prove that each counting function of the $m$-simple branched covers 
with a fixed genus of an elliptic curve 
is expressed as a polynomial of the Eisenstein series
$E_2$, $E_4$ and $E_6$.
The special case $m=2$ is considered by Dijkgraaf.

\section{Introduction}

We consider the counting function
\[
F^{(m)}_g(q) = \sum_{d\ge 1} N_{g,d}^{(m)} q^d
\]
of the branched covers of
an elliptic curve.
Here, $N_{g,d}^{(m)}$ is 
the (weighted) number of 
isomorphism classes of branched covers,
with genus $g(>1)$, degree $d$,
and ramification index $(m,m,\dots,m)$,
of an elliptic curve.
Such a cover is called an $m$-simple cover.
Our aim is to prove that
the formal power series $F^{(m)}_g$ 
converges to a function
belonging to the graded ring 
of  {\it quasi}-modular forms
with respect to the full modular group $SL(2,\Z)$,
and indeed that it can be expressed 
as a polynomial of the Eisenstein series
$E_2$, $E_4$ and $E_6$ with rational coefficients.

For $m=2$, an $m$-simple branched cover is
a simple branched cover.
Dijkgraaf \cite{Dijkgraaf} has proved that
the counting function $F^{(2)}_g(q)$ 
is a quasi-modular form with respect to $SL(2,\Z)$.
Our result is a generalization of this result
for arbitrary $m \ge 2$.

The proof \cite{Dijkgraaf} for $m=2$ employs the `Fermionic formula'
\cite{Douglas} of the partition function,
\begin{eqnarray*}
&& \exp\left(
\sum_{g=1}^\infty F_g^{(2)}(q) \frac{X^{2g-2}}{(2g-2)!} \right) \\
&=& q^{-1/24}
\displaystyle\Res_{z=0} \left(\prod_{p \in \frac12+\Z_{\ge 0}} 
(1 + z      q^p \exp( p^2 X/2))
(1 + z^{-1} q^p \exp(-p^2 X/2)) \frac{dz}{z} \right),
\end{eqnarray*}
whose quasi-modularity was proven by Kaneko and Zagier \cite{KZ}.
The quasi-modularity of the counting function $F_g^{(2)}$
supports the mirror symmetry for an elliptic curve.
For $m\ge 3$,
although the relation between the counting function
$F_g^{(m)}$ and the theory of mirror symmetry 
has not yet been clarified,
the quasi-modularity of the counting function 
has been shown to hold.

The proof of our main theorem, Theorem~\ref{thm:main}, 
implies that all counting functions $F_q^{(m)}$ with $m\ge 2$ and
$g > 1$ {\it live in} the infinite product
\begin{eqnarray*}
&& V(q,t_2,t_3,\dots) 
= \exp({- \sum_{j=1}^\infty \xi(-j) t_j}) \times  \\
&& \qquad \displaystyle\Res_{z=0} \left(\prod_{p \in \frac12+\Z_{\ge 0}} 
(1 + z      q^p \exp( \sum_{k\ge2}    p^k t_k))
(1 + z^{-1} q^p \exp(-\sum_{k\ge2} (-p)^k t_k)) \frac{dz}{z} \right),
\end{eqnarray*}
with the infinite set of variables $q,t_2,t_3,\dots$,
where the renormalizing factor $\xi(-j)$ is
the special value of a Hurwitz zeta function.
To be more precise, 
we show that every $F_g^{(m)}$ 
is a linear combination of the Taylor coefficients
of the function $V$.
Then, the quasi-modularity of the counting function $F_g^{(m)}$ 
is derived from the corresponding property
for $V$, which was established by
Bloch-Okounkov \cite{BO}.
The key step in the proof is Proposition~\ref{th:phi}.
In fact,
the counting function is expressed 
as a combinatorial sum 
related to the symmetric group $S_d$.
This expression depends strongly on 
the size $d$ of the symmetric group $S_d$.
To complete the summation in $d$,
we need some formula free of $d$.
Such a formula is obtained in Proposition~\ref{th:phi}.

The author expresses his gratitude to Professor Masanobu Kaneko
for helpful discussions.

\section{Counting functions}
\subsection{$m$-simple branched cover}

We fix an elliptic curve $E$ over $\C$
and an integer $m\ge 2$.
A pair $(f,C)$ consisting of a (smooth complex) curve $C$ and 
a holomorphic map $f:C \rightarrow E$ is
{\it an $m$-simple branched cover} 
if the following three conditions are safisfied:
\begin{enumerate}
\item[(i)] $C$ is connected.
\item[(ii)] For any $P\in C$, the branching index $e(P)=1$ or $m$.
\item[(iii)] If $P \neq P'$ and $e(P)=e(P')=m$, then $f(P) \neq f(P')$.
\end{enumerate}
In the case $m=2$, a $2$-simple branched cover is usually called 
a `simple branched cover'. 
An $m$-simple branched cover is a natural generalization of
a simple branched cover.
If $f$ is of degree $d$ and the curve $C$ is of genus $g$, 
then the pair $(f,C)$ is said to be of genus $g$ and degree $d$.

Two $m$-simple branched covers $(f,C)$ and $(f',C')$ are
isomorphic if
there is an isomorphism $\varphi:C\rightarrow C'$ such that
$f=f'\circ\varphi$.
The group of automorphisms on $(f,C)$ is denoted 
$\Aut(f,C)$ [or simply $\Aut(f)$]. 
We will see that this is a finite group.

By the Riemann-Hurwitz formula (e.g., \cite{GH}), we have
\[
2g(C) -2 = d(2 g(E)-2) + \sum_{P \in C} (e(P)-1).
\]
Thus the number $b$ of branch points 
and the genus $g$ of the curve $C$ always satisfy
the relation $2g-2= (m-1)b$.
Note that the genus $g$ does not depend on the degree $d$.
This relation implies that
the number $b$ of branch points should be even if $m$ is even.
If $m$ is odd, the number of branch points is arbitrary.
The case $g=1$ corresponds to the case $b=0$;
that is, the cover $f:C\rightarrow E$ is unramified.

We choose $b$ (distinct) points $P_1,\dots,P_b \in E$.
For $g=1+(m-1)b/2$,
let $X_{g,d}=X_{g,d}^{(m)}$ be the set of isomorphism classes 
of $m$-simple branched covers of genus $g$ and degree $d$
such that $e(P_i)=m$ for $i=1,\dots,b$,
and $e(P') =1$ for $P' \in E\setminus \{ P_1,\dots,P_b \}$.
We will see that
$X_{g,d}$ is a finite set
and does not depend on the choice of the set of branch points
$P_1,\dots, P_b$.
In fact, $X_{g,d}$ can also be regarded as the fiber in the fibration
\[
X_{g,d} \rightarrow \mathcal{M}_g(E,d) \rightarrow E_{b},
\]
where 
$\mathcal{M}_g(E,d)$ is the Hurwitz space of $m$-simple branched covers, 
and $E_b$ is the configuration space of unordered $b$-points on $E$.

We count the (weighted) number of elements of $X_{g,d}$ so that
\[
N_{g,d} = \sum_{f \in X_{g,d}} 
\frac{1}{|\Aut(f)|}.
\]
Note that $N_{g,d}=0$ unless $2(g-1) \in (m-1) \Z_{\ge0}$.
The generating functions $F_g$ for $g>1$ are now defined by
\[
F_g(q) = F_g^{(m)}(q) = \sum_{d\ge1} N_{g,d} q^d.
\]
These functions are called the `counting functions'.

It is necessary to define $F_1$ separately.
The case $g=1$ must be treated separately
because 
covers in this case are unbranched ($b=0$).
Note that neither $X_{1,d}$ or $N_{1,d}$ depends on $m$.
Then we employ the definition of $F_1(q)$ 
introduced for the case $m=2$ \cite[\S2]{Dijkgraaf}:
\[
F_1(q) = - \frac1{24} \log q + \sum_{d\ge 1} N_{1,d} q^d.
\]
Here, the first term can be considered
the contribution of the constant map (the map of degree zero)
which is not a stable map.
Since $N_{1,d} = \sigma_1(d)/d$, where $\sigma_1(d)$ is the sum of
all divisors of $d$,
we have the expression
\[
F_1(q) = - \log \eta(q),
\]
where we denote the Dedekind eta function by
\[
\eta(q) = q^{1/24} \prod_{n\ge1} (1-q^n).
\]

Next, we introduce  a two-variable partition function $Z$,
\begin{eqnarray*}\label{eqn:def Z}
Z(q,\vari) = Z^{(m)}(q,\vari) 
&:= &  \exp\left(\sum_{g\ge 1}  F_g(q)  
   \displaystyle\frac{\vari^{(2g-2)/(m-1)}}{((2g-2)/(m-1))!} 
         \right) \\
& = &  \exp\left(\sum_{b\ge 0}  F_{1+(m-1)b/2}(q)  
   \displaystyle\frac{\vari^{b}}{b!} \right),
\end{eqnarray*}
which is a formal power series in $q$.
We see that
\begin{eqnarray}\label{eqn:eta Z}
\eta(q) Z(q,\vari) &=& 
\exp\left(\sum_{g\ge2}  F_g(q)  
   \displaystyle\frac{\vari^{(2g-2)/(m-1)}}{((2g-2)/(m-1))!} \right) \\
&=& 
\exp\left(\sum_{b\ge1}  F_{1+(m-1)b/2}(q)  
   \displaystyle\frac{\vari^{b}}{b!} \right).
\end{eqnarray}

In the definition of the counting function $F_g$,
 we consider only connected covers.
We also introduce 
the partition function $\hat{Z}$ of
the counting functions of covers, which are not necessarily connected.
Let $\hat{X}_{g,d}$ 
be the set of isomorphism classes of $m$-simple branched covers,
which are not necessarily connected, of genus $g$ and degree $d$.
In other words, for $\hat{X}_{g,d}$, 
we impose conditions~(ii) and (iii),
but we drop condition~(i).
We define the corresponding (weighted) number of elements of $\hat{X}_{g,d}$
by
\[
\hat{N}_{g,d} = \sum_{f \in \hat{X}_{g,d}} 
\frac{1}{|\Aut(f)|},
\]
the modified counting function $\hat{F}_g$ for $g\ge1$ by
\[
\hat{F}_g(q) = \sum_{d\ge1} \hat{N}_{g,d} q^d,
\]
and its generating function $\hat{Z}$ by
\begin{eqnarray*}
\hat{Z}(q,\vari) 
&=& \sum_{g \ge 1} \hat{F}_g(q)  
  \displaystyle\frac{\vari^{(2g-2)/(m-1)}}{ ((2g-2)/(m-1))! } \\
&=& \sum_{b\ge 0} \hat{F}_{1+(m-1)b/2}(q)
  \displaystyle\frac{ \vari^{b} }{ b! }.
\end{eqnarray*}
The relation between these two functions is given as follows.
\begin{lemma}\label{th:Zhat}
We have the relation $\hat{Z}(q,\vari) = q^{1/24} Z(q,\vari)$.
\end{lemma}
Proof: 
This follows from a standard argument
\cite{Dijkgraaf}.
\qed

\subsection{Representations of the fundamental group}
The weighted number $\hat{N}_{g,d}$ of
covers which are not necessarily connected
is expressed in terms of representations of the fundamental group
of the punctured elliptic curve.

Let $\pi_1^b$ be the fundamental group
of the $b$-punctured curve $E \setminus \{ P_1, \dots, P_b \}$.
It is known that the group $\pi_1^b$ can be expressed
in terms of the generators and relations as
\[
\pi_1^b=\langle \alpha,\beta,\gamma_1,\dots, \gamma_b \mid 
\gamma_1\cdots\gamma_b = \alpha\beta\alpha^{-1}\beta^{-1} \rangle.
\]
Here, we denote the simple curve around a point $P_i$ by
$\gamma_i \in \pi_1(E')$.

Let $S_d$ be the symmetric group $S_d$ on $d$ elements,
and let $c^{(m)}$ be the conjugacy class of $S_d$ of type $(m,1^{d-m})$.
In other words, the class $c^{(m)}$ consists of  cycles of length $m$.
We define
\[
\Phi_{g,d} = \Phi^{(m)}_{g,d}
= \{ \varphi \in \Hom(\pi_1^b, S_d) 
\mid \varphi(\gamma_i) \in c^{(m)} \mbox{ for } i=1,\dots,b \},
\]
where ``$\Hom$'' represents the set of group homomorphisms.
The symmetric group $S_d$ acts on $\Phi_{g,d}$ 
according to 
\[
\varphi^\sigma(\gamma) = \sigma^{-1} \varphi(\gamma) \sigma,
\qquad 
\sigma \in S_d, \varphi \in \Phi_{g,d}.
\]

\begin{lemma}\label{th:Nhat}
\begin{enumerate}
\item[{\rm (i)}]
As a set, we have the bijection
 $\hat{X}_{g,d} \cong \Phi_{g,d}/S_d $.
\item[{\rm (ii)}]
$ \hat{N}_{g,d} = | \Phi_{g,d} | / |S_d| . $
\end{enumerate}
\end{lemma}
Proof:
(i)
Let 
$E' = E\setminus \{ P_1 ,\dots, P_b \}$ be 
a punctured curve.
Let us choose a base point $P_0 \in E'$ 
as a base point.
Then the fundamental group $\pi_1(E') = \pi_1(E', P_0)$
is isomorphic to $\pi_1^b$.
For an $f \in \hat{X}_{g,d}$,
we construct the corresponding map $\varphi \in \Phi_{g,d}$.
Let 
$f^{-1}(P_0) = \{ Q_1, \dots, Q_d \}$.
Then we have the natural map
\[
\varphi: \pi_1^b \cong \pi_1(E') \rightarrow \Aut(f^{-1}(P_0)) \cong S_d.
\]
Conversely,
for each $\varphi \in \Phi_{g,d}$,
we construct a covering $f \in \hat{X}_{g,d}$.
We denote the universal covering of $E'$ by
$E'{}^{univ}$.
Let $C' = E'{}^{univ} \times_\varphi \{ 1,\dots,d\}
= E' \times \{1, \dots, d\} / \sim$,
where $(x,i) \sim (\gamma x, \varphi(\gamma) i)$
when $\gamma \in \pi_1(E')$, $x \in E'{}^{univ}$ and $1 \le i \le d$.
Then the natural projection 
$f': C' \rightarrow E'{}^{univ} / \pi_1^b = E'$
is a covering of degree $d$.
This extends to a ramified covering $f:C \rightarrow E$.
It is easy to see that this construction gives
the required bijection.

(ii) Under the bijection in (i),
the group $\Aut(f)$ of automorphisms corresponds to 
the stabilizer subgroup of $S_d$ at $\varphi$.
This implies that
\[
| \Aut(f) | = \# \{ \sigma \in S_d \mid \varphi = \varphi^\sigma \}.
\]
Then we have
\[
\hat{N}_{g,d} = \sum_{f} \frac{1}{|\Aut(f)|}
= \frac{1}{|S_d|} \sum_{f} \# \{ \varphi^\sigma \mid \sigma \in S_d,
\varphi \mbox{ corresponds to } f \}
= |\Phi_{g,d}|/|S_d|.
\]
\qed

\subsection{Irreducible characters of symmetric group}
The number of group homomorphisms appearing in the previous lemma
is written as a sum over the irreducible representations
of the permutation group.

A partition $\parti=(\parti_1,\parti_2,\dots)$ of $d$
is a non-increasing sequence 
$
\parti_1 \ge \parti_2 \ge \dots
$
of non-negative integers such that $\sum_{i=1}^d \parti_i =d$.
We denote the set of all partitions of  $d$ by $\Young$.
It is known that
the set of irreducible representations of the
symmetric group $S_d$ is
parametrized by $\Young$.
For each $\parti \in \Young$, 
we denote the corresponding irreducible character by $\chi_\parti$.
Since a character is a class function,
the value $\chi_\parti(c)$ is well-defined 
for each conjugacy class $c$ of $S_d$.
We introduce the modified character 
\[
f_\parti(c) = \frac{|c|\cdot \chi_\parti(c)}{\dim \parti},
\]
where $|c|$ is the number of elements in the conjugacy class $c$,
and $\dim \parti$ is the dimension of the representation $\parti$,
that is, the value of $\chi_\parti(e)$ at the identity of $S_d$.

\begin{lemma}\label{th:Phi}
For $g=1+(m-1)b/2$, we have
\[
|\Phi_{g,d}^{(m)}| / |S_d| = \sum_{\parti \in \Young} f_\parti(c^{(m)})^b.
\]
\end{lemma}
Proof : We apply the formula in Lemma~4 of \cite{Dijkgraaf}
with 
$G=S_d$, $R=\Young$, $c_1=\cdots=c_N=c^{(m)}$, $h=1$ and $N=b$.
\qed
\subsection{Frobenius notation}
Now we recall properties of Frobenius coordinates of partitions
and shifted symmetric functions.
Our Frobenius coordinates are parametrized by half-integers,
not by integers, as is explained below.

For a partition $\parti=(\parti_1,\dots,\parti_d) \in \Young$,
we define the shifted partition
$\sparti=(\sparti_1,\dots,\sparti_d)$ by
$\sparti_i=\parti_i-i+\frac12$.
Let $I$ be the set of positive half-integers, 
$I=\frac12+\Z_{\ge0} = \{ \frac12, \frac32,\dots\}$.
A partition $\parti$ gives us two subsets $P,Q \subset I$ such that
\begin{eqnarray*}
P &=& \{ \sparti_i  \mid \sparti_i > 0,  i=1,\dots,d \}, \\
Q &=& \{ 1/2, 3/2, \dots, (2d-1)/2 \}
    \setminus  \{ - \sparti_i  \mid - \sparti_i> 0, i=1,\dots,d \}
 = \{ \tilde{\parti'}_i \mid \tilde{\parti'}_i, i=1, \dots, d \},
\end{eqnarray*}
where $\parti'$ is the conjugate partition of $\parti$.
Then the cardinality of $P$ equals that of $Q$.
Conversely, for a given pair of subsets $P, Q \subset I$
with $|P|=|Q|$, 
we have the corresponding the partition $\parti \in \Young$
with $d = |P|+|Q|= 2|P|$.

We remark that our Frobenius coordinates $(P,Q)$
are shifted by $1/2$ from the Frobenius coordinates
$(\alpha_1,\dots, \alpha_r \mid \beta_1,\dots,\beta_r)$
introduced in Section~I.1 of \cite{Macdonald}.
The precise relation is 
\[
P = \{ \alpha_1+ \frac12, \alpha_2+ \frac12, \dots, \alpha_r+\frac12 \},
\qquad
Q = \{ \beta_1 + \frac12, \beta_2 +\frac12, \dots, \beta_r + \frac12 \}.
\]

For $k \in \Z_{\ge0}$ we define
\[
\ps{k}{\parti} = 
\sum_{i=1}^d \left(\sparti_i^k - (-i+\frac12)^k \right).
\]
(This function is written as $p_k(\parti)$ in (5.4) of \cite{BO}.
For example,
$\ps{0}{\parti} = 0$,
$\ps{1}{\parti} = d$.
  From I.1.4 of \cite{Macdonald} we have the relation
\[
\sum_{i=1}^d (t^{\sparti_i} - t^{-i+\frac12})
= \sum_{p \in P} t^p - \sum_{p \in Q} t^{-p},
\]
where $(P,Q)$ is the Frobenius coordinates of the partition $\lambda$.
As a corollary, we have  \cite[(5.4)]{BO}
\[
\ps{k}{\parti}
= \sum_{i=1}^d \left( \sparti_i^k - (-i+\frac12)^k \right)
= \sum_{p \in P} p^k - \sum_{p \in Q} (-p)^k.
\]
This is a power-sum symmetric functions in 
$\sparti=(\sparti_1,\dots,\sparti_d)$
plus some polynomial in $d$ of degree $k+1$.
We now introduce two additional polynomials symmetric in the $\sparti_i$.
Let $e_j(\sparti) $
be the $j$th elementary symmetric function
and $h_j(\sparti)$
 the $j$th complete symmetric function,
defined by
\begin{eqnarray*}
e_j(\sparti) 
&=& \sum_{1\le i_1<\cdots < i_j \le d} 
\sparti_{i_1} \cdots \sparti_{i_j}, \\
h_j(\sparti)
&=& \sum_{1\le i_1 \le \cdots \le i_j \le d} 
\sparti_{i_1} \cdots \sparti_{i_j}. 
\end{eqnarray*}
These two functions can be expressed as polynomials in
power-sum symmetric functions,
and thus as polynomials in $\ps{k}{\parti}$ and $d$.
\subsection{Character formula}
The character value $f_\parti(c^{(m)})$ can be written 
in terms of $\ps{k}{\parti}$.
Although the character depends strongly on
the rank $d$ of the symmetric group $S_d$,
the following expression is independent of $d$.
It is thus useful for summation over $d$.
\begin{proposition}\label{th:phi}
There exists a polynomial 
$\phi_m(Y_1,\dots,Y_m) \in \Q[Y_1,\dots,Y_m]$
such that for all  $d\ge1$ and $\parti \in \Young$, we have
\[
f_\parti(c^{(m)}) = \phi_m(\ps{1}{\parti}, \dots, \ps{m}{\parti}).
\]
\end{proposition}
Proof:
We consider a partition $\parti=(\parti_1,\dots,\parti_d)$.
Let
\[
\mu_i = \parti_i + d-i = \sparti_i + d -\frac12,
\]
and $\varphi(x) = \displaystyle\prod_{i=1}^d (x-\mu_i)$.
Then, from  Example~I.7.7 in \cite{Macdonald}, we have
\[
f_\parti(c^{(m)}) =
\frac{1}{m^2} \Res_{x=\infty} 
\left( \frac{x(x-1)\cdots(x-m+1) \varphi(x-m)}{\varphi(x)} dx \right),
\]
where the expression ``$\Res_{}$'' denotes the residue.
Since 
$
\varphi(x+d-\frac12) = \prod_{i=1}^d (x-\sparti_i),
$
we obtain
\begin{eqnarray*}
& & f_\parti(c^{(m)}) \\
&=&
\frac{1}{m^2} \Res_{x=\infty} 
\left((x+d-\frac12)(x+d-\frac32)\cdots(x+d-m+\frac12) 
\frac{\varphi(x-m+d-\frac12)}{\varphi(x+d-\frac12)} dx \right) \\ 
& = & - 
\frac{1}{m^2} \Res_{y=0} \left(
(1+(d-\frac12)y)(1+(d-\frac32)y)\cdots(1+(d-m+\frac12)y) 
\frac{\prod_{i=1}^d (1-(m+\sparti_i)y)}{\prod_{i=1}^d (1-\sparti_i y)}
\frac{dy}{y^{m+2}} \right)
\end{eqnarray*}
by changing coordinates.
The products appearing here are generating functions
of elementary (resp. complete) symmetric functions:
\begin{eqnarray*}
\prod_{i=1}^d (1-(m+\sparti_i)y)
& = & \displaystyle\sum_{j=0}^d (1-my)^{d-j}  (-y)^j e_j(\sparti), \\
\displaystyle \prod_{i=1}^d (1-\sparti_i y)^{-1}
& = & \displaystyle\sum_{j=0}^\infty y^j h_j(\sparti),
\end{eqnarray*}
Then,
\begin{eqnarray*}
&& f_\parti(c^{(m)}) \\
&=& - \frac{1}{m^2}
\displaystyle \sum_{i=0}^d \sum_{j=0}^\infty e_i(\sparti) h_j(\sparti) \times \\
&&  \qquad
\displaystyle \Res_{y=0}
\left(
(1+(d-\frac12)y)(1+(d-\frac32)y)\cdots(1+(d-m+\frac12)y) 
(1-my)^{d-i} (-y)^i y^j \frac{dy}{y^{m+2}} \right) \\
& =&  - \frac{1}{m^2} 
\displaystyle{\sum_{i=0}^d \sum_{j=0}^\infty (-1)^i e_i(\sparti) h_j(\sparti) }
b_{ij},
\end{eqnarray*}
where
\[
b_{ij} = 
\Res_{y=0} 
\left(
(1+(d-\frac12)y)(1+(d-\frac32)y)\cdots(1+(d-m+\frac12)y) 
(1-my)^{d-i} \frac{dy}{y^{m+2-i-j}} \right).
\]

\begin{lemma}\label{lemma:bij}
\begin{enumerate}
\item[{\rm (i)}]
$b_{ij}=0$ for $i+j \ge m+2$.
\item[{\rm (ii)}]
For $i+j = m+1$,
we have $b_{ij}= 1$ and 
\[
\sum_{i+j=m+1} (-1)^i e_i(\sparti) h_j(\sparti) b_{ij} = 0.
\]
\item[{\rm (iii)}]
For $i+j=m$, we have $b_{ij} = \frac{m^2}{2} + mi$.
\item[{\rm (iv)}]
For $0 \le i+j \le m$,
the value $b_{ij}$ depends on $d$ polynomially.
In fact, $b_{ij}$ is a polynomial in $d$ of degree 
$m+1-i-j$ with coefficients in $\Q$.
\end{enumerate}
\end{lemma}
Proof: (i)
If $i+j \ge m+2$, then the function inside the summation
is a polynomial in $y$,
and thus it has no pole at $y=0$ and its residue $b_{ij}$ is $0$.

(ii) 
If $i+j=m+1$, then the residue $b_{ij}$ is $1$, and 
the contribution to this sum is, as in I.2.$6'$ of \cite{Macdonald},
\[
\sum_{i+j=m+1} (-1)^i e_i(\sparti) h_j(\sparti) = 0,
\]
which is the coefficient of $y^{m+1}$ in
$\prod_{i=1}^d (1-\sparti_i y) / \prod_{i=1}^d (1-\sparti_i y)=1$.

(iv)
Since $b_{ij}$ is the coefficient
of $y^{m+1-i-j}$ in the polynomial
\begin{eqnarray*}
&&
(1+(d-\frac12)y)(1+(d-\frac32)y)\cdots(1+(d-m+\frac12)y) 
 (1-my)^{d-i} \\
&=&
\sum_{s=0}^{\infty} \sum_{t=0}^m
e_t(d-\frac12,d-\frac32,\dots,d-m+\frac12) \binom{d-i}{s} (-m)^s y^{s+t}, \\
\end{eqnarray*}
we have
\[
b_{ij} = 
\sum_{s=0}^{m+1-i-j} 
e_{m+1-i-j-s}(d-\frac12,d-\frac32,\dots,d-m+\frac12) 
  \binom{d-i}{s} (-m)^s.
\]
Then $b_{ij}$ is a polynomial in $d$ of degree no greater than $m+1-i-j$.

(iii)
If $i+j=m$, then
\[
b_{ij}
=
(d-\frac12)+(d-\frac32)+\cdots+(d-m+\frac12)-m(d-i)
=\frac{m^2}2 + m i.
\]
This implies that
\[
-\frac{1}{m^2} \sum_{i+j=m} (-1)^i e_i(\sparti) h_j(\sparti) b_{ij}
= -\frac1m \sum i (-1)^i e_i(\sparti) h_{m-i}(\sparti)
= \frac1m \left( \ps{m}{\parti} + \sum_{i=1}^d (-i+\frac12)^m \right).
\]
\qed
We now return to the proof of Proposition~\ref{th:phi}.
We have the finite sum expression
\[
 f_\parti(c^{(m)}) 
= - \frac{1}{m^2} 
  \sum_{i+j\le m} (-1)^i e_i(\sparti) h_j(\sparti) b_{ij}.
\]
This is a polynomial in $e_i$, $h_j$ and $d$.
We know that $e_i$ and $f_j$ are
polynomials in power-sum symmetric functions $\ps{k}{\parti}$ and $d$.
Then, since $d= \ps1{\lambda}$,
we have proved the existence of the function $\phi=\phi_m$.
\qed   
\begin{example}
For $m=2,\dots,5$, 
the polynomial $\phi_m$ is of the following form:
\[
\phi_2=\frac12 Y_2, \quad
\phi_3=\frac13 Y_3 -\frac12 Y_1^2 + \frac{5}{12} Y_1, \quad
\phi_4=\frac14 Y_4 - Y_1 Y_2 + \frac{11}{8} Y_2,
\]
\[
\phi_5=\frac15 Y_5 - Y_3 Y_1 + \frac{19}{6} Y_3 - \frac12 Y_2^2
 + \frac56 Y_1^3 -\frac{15}4 Y_1^2 + \frac{189}{80} Y_1.
\]
\end{example}
This suggests that
the degree of the polynomial $\phi_m$ would be $m$
if we consider the degree of $Y_j$ to be $j$.
The highest order term of $\phi_m$ 
would then be $\displaystyle\frac{1}{m}Y_m$.

For $m=2$,
$\ps2\parti /2 = f_\parti(c^{(2)})$ has a simple expression 
in terms of partitions.
For a partition $\parti$, we define
$n(\parti)=\sum_{i\ge 1} (i-1) \parti_i$.
We also define 
the content $c(x)$ as
$c(x) = j-i$ for each box $x =(i,j) \in \parti$.
Then
\[
\ps2\parti /2 = f_\parti(c^{(2)}) = n(\parti') - n(\parti) 
= \sum_{x \in \parti} c(x).
\]


\section{Quasi-modular form}
\subsection{Eisenstein series}
We give a brief summary of quasi-modular forms
to fix the notation used here.
(For the precise definition and further properties,
see \cite{KZ} and \S3 of \cite{BO}.)
Let $\tau$ be a complex number with $\Im\tau>0$
and $q=e^{2\pi \sqrt{-1} \tau}$.
We denote the differential operator 
$\frac{1}{2\pi \sqrt{-1}} \frac{d}{d\tau} = q \frac{d}{dq} $
by $\Dq$.
For a subgroup $\Gamma$ of the full modular group $SL(2,\Z)$ of finite index,
we denote the set of modular forms of weight $k$ by
$M_k(\Gamma)$ and the graded ring of modular forms
by $M_*(\Gamma) = \displaystyle\oplus_{k\ge 0} M_k(\Gamma)$.
Similarly,
we denote the set of quasi-modular forms of weight $k$ by
$\QM_k(\Gamma)$ and the graded ring of quasi-modular forms
by $\QM_*(\Gamma) = \displaystyle\oplus_{k\ge 0} \QM_k(\Gamma)$.
The ring $M_*(\Gamma)$ is not closed under the 
differentiation $\Dq$, but the
ring  $\QM_*(\Gamma)$ is closed under $\Dq$.
Examples of (quasi-)modular forms are provided by 
the Eisenstein series.

We denote the Bernoulli number by $B_k \in \Q$,
which is defined by
$\displaystyle\frac{x}{e^x - 1} = \sum_{k=0}^\infty B_k \frac{x^k}{k!}$.
For example,
$B_0 =1$, $B_1 = -\frac12$, $B_2 = \frac16$, $B_4= -\frac1{30}$ and
$B_6 = \frac1{42}$.

We define the (normalized) Eisenstein series $E_k$ for even $k \ge 4$ by
\begin{eqnarray*}
E_k(\tau) 
&=& \frac{1}{2} \sum_{(c,d)=1} \frac{1}{(c \tau + d)^k} \\
&=& 1 - \frac{2k}{B_k} \sum_{n=1}^\infty
(\sum_{d|n} d^{k-1}) q^n
= 1 - \frac{2k}{B_k} \sum_{n=1}^\infty \frac{n^{k-1} q^n}{1-q^n}.
\end{eqnarray*}
(This is a convergent series in $q$.)
Then $E_k$ is a modular form of weight $k$ for $SL(2,\Z)$:
\[
E_k(\frac{a \tau + b}{c \tau + d}) = (c\tau + d)^k E_k(\tau).
\]
We also define
\[
E_2(\tau) = 1 - 24 \sum_{n=1}^\infty (\sum_{d|n} d) q^n.
\]
Then $E_2$ is not a modular form, 
but a quasi-modular form of weight $2$ for $SL(2,\Z)$, so that
\[
E_2(\frac{a \tau + b}{c \tau + d}) = (c\tau + d)^2 E_2(\tau)
+ \frac{12}{2\pi \sqrt{-1}} c (c\tau +d).
\]

\begin{lemma}
\begin{enumerate}
\item[{\rm (i)}] 
$\QM_*(\Gamma)$ is a graded $\C$-algebra,
 and $M_*(\Gamma) \subset \QM_*(\Gamma)$ is a subalgebra.
\item[{\rm (ii)}]
 As a $\C$-algebra, we have the isomorphism
$ \QM_*(\Gamma) = M_*(\Gamma) \otimes \C[E_2] $.
\item[{\rm (iii)}]
 $\QM_*(\Gamma)$ is stable under the action of $\Dq$.
(It increases the degree by $2$.)
\end{enumerate}
\end{lemma}

For the full modular group, 
since  $M_*(SL(2,\Z)) = \C[ E_4, E_6 ]$,
we have $\QM_*(SL(2,\Z)) = \C[ E_2, E_4, E_6] $.
The differentiation $\Dq$ provides the dynamical system
\[
\Dq(E_2) = (E_2^2-E_4)/12, \quad
\Dq(E_4) = (E_2 E_4 -E_6)/3, \quad
\Dq(E_6) = (E_2 E_6 - E_4^2)/2.
\]

The following lemma is used for the proof
of the main theorem.
\begin{lemma}\label{lem:eta D}
If $\eta(q) A(q) \in \QM_k(SL(2,\Z))$,
then $\eta(q) \Dq^j (A(q)) \in  \QM_{k+2j}(SL(2,\Z))$
for a positive integer $j$.
\end{lemma}
Proof:
Recall the definition of the Ramanujan delta,
$\Delta(\tau) = \eta(q)^{24} = (E_4^3 - E_6^2)/1728$.
Then we have
$ \Dq \log \Delta(\tau) = E_2(\tau) $ and
$ \Dq (\log \eta) = E_2/24$,
and we obtain the formula
\[
\eta(q) \Dq A(q) = \Dq(\eta(q) A(q)) - \frac1{24} E_2 \eta(q) A(q).
\]
The condition $\eta(q) A(q) \in \QM_k(SL(2,\Z))$ implies
$\eta(q) \Dq (A(q)) \in \QM_{k+2}(SL(2,\Z))$.
The assertion follows from induction on $j$.
\qed
\subsection{The character of the infinite wedge representation}

We introduce the variables $t_1, t_2, t_3,\dots$,
where we write $D_k = \frac{\partial}{\partial t_k}$ for $k\ge 1$.
In what follows, the variable $t_1$ is related to $q$
by $q=e^{t_1}$.
In particular, for $k=1$ we have
$D=D_1 = q \frac{\partial}{\partial q}$.
We define the infinite series 
\begin{eqnarray}\label{eqn:V'}
V'(q,t_2,t_3,\dots)
& = & \displaystyle\sum_{d\ge0} \sum_{\parti \in \Young}
\exp(\ps{1}{\parti} t_1
   + \ps{2}{\parti} t_2
   + \ps{3}{\parti} t_3+ \cdots) \\
& = & \displaystyle\sum_{d\ge0} \sum_{\parti \in \Young}
        q^{\ps{1}{\parti}}
  \exp(\ps{2}{\parti} t_2
     + \ps{3}{\parti} t_3 + \cdots).
\end{eqnarray}
This expression appears in (0.10) of \cite{BO}
as a character of the infinite wedge representation of
an infinite dimensional Lie algebra ($W_\infty$),
and it is known to be a quasimodular form of weight $-\frac12$.
Let us explain this in more detail.

It is easy to see that $V'$ is the coefficient of $z^0$
of an infinite product:
\begin{eqnarray*}
V'
& = &
\displaystyle\Res_{z=0} \left(\prod_{p \in \frac12+\Z_{\ge 0}} 
(1 + z      \exp( \sum_{k\ge1}    p^k t_k))
(1 + z^{-1} \exp(-\sum_{k\ge1} (-p)^k t_k)) \frac{dz}{z} \right) \\
& = &
\displaystyle\Res_{z=0} \left(\prod_{p \in \frac12+\Z_{\ge 0}} 
(1 + z      q^p \exp( \sum_{k\ge2}    p^k t_k))
(1 + z^{-1} q^p \exp(-\sum_{k\ge2} (-p)^k t_k)) \frac{dz}{z} \right).
\end{eqnarray*}
To obtain a quasimodular form,
we have to multiply a fractional power in $t_i$. 
Let $\xi(s) =  \sum_{n\ge1} (n-\frac12)^{-s}
= (2^s-1)\zeta(s)$,
which is continued to a meromorphic function of $s$.
The function $\xi(s)$ at negative integer values of $s$ 
is well-defined,
and $\xi(-2i)=0$ for $i \in \Z_{>0}$.
(For example, $\xi(-1) = 1/24$, $\xi(-3) = -7/960$.)
We define
\begin{equation}\label{eqn:VV'}
V(q,t_2,\dots) 
= \exp({- \sum_{j=1}^\infty \xi(-j) t_j}) \times  V'(q,t_2,\dots).
\end{equation}
If we consider the case $t_2=t_3=\cdots=0$, then
the infinite product reduces to
\[
q^{-\xi(-1)}
\prod_{p \in \frac12+\Z_{\ge 0}} 
(1+z q^p )(1+z^{-1} q^p )
= \frac{\sum_{n\in\Z} z^n q^{n^2/2}}{\eta(q)}
\]
since $\xi(-1)= 1/{24}$.
Then
\begin{equation}\label{eqn:eta V0}
\eta(q) V(q,0,0,\dots) = 1.
\end{equation}
Now,  consider the Taylor expansion of $V$ with respect to
$(t_2,t_3,\dots)$ 
\begin{equation}\label{eqn:Taylor}
V(q,t_2,t_3,\dots)= \sum_{K} A_K(q) \frac{t^K}{K!},
\end{equation}
where $K=(k_2,k_3,\dots)$ with almost all $k_i=0$,
and $t^K/K! =t_2^{k_2} t_3^{k_3} \cdots / k_2! k_3! \cdots$
is multi-index notation.
The relation (\ref{eqn:eta V0}) implies that 
$\eta(q) A_{(0,0,\dots)}(q) = 1$.
It is known in (4.8) of \cite{BO} that
$\eta(q) A_K(q) \in \QM_*(SL(2,\Z))$
and is 
of weight $3k_2 + 4 k_3 + \cdots = \sum_{i=2}^\infty (i+1) k_i$.
By Lemma~\ref{lem:eta D}, 
we know that $\eta(q) \Dq^j (A_K(q)) \in \QM_*(SL(2,\Z))$
and its weight is $2j + \sum_{i=2}^\infty (i+1) k_i$.

\subsection{Main theorem}
We arrive at the stage to state our main theorem.

\begin{theorem}\label{thm:main}
The counting functions $F_g(q)= F_g^{(m)}(q)$ 
for $g\ge2$ belong to the graded ring $\QM_*(SL(2,\Z))$
of quasimodular forms 
with respect to the full modular group $SL(2,\Z)$.
In particular, $F_g^{(m)} \in \Q[E_2,E_4,E_6]$.
\end{theorem}
Proof: 
Summarizing Lemmas~\ref{th:Zhat}, \ref{th:Nhat} and
\ref{th:Phi}, we obtain 
\begin{equation}\label{shiki:Zhat}
\hat{Z}(q,\vari) 
= 1+\sum_{b \ge 0} \sum_{d \ge 1} \sum_{\parti \in \Young}
\frac{1}{b!}  f_\parti(c^{(m)})^b q^d \vari^{b}
= 1+\sum_{d \ge 1} \sum_{\parti \in \Young} \exp(f_\parti(c^{(m)}) \vari) q^d.
\end{equation}
We can consider 
the term $1$ as coming from the case $d=0$, 
where $R_0 = \{ \emptyset \}$, $f_{\emptyset} = 0$.
   From Proposition~\ref{th:phi},
we obtain
\begin{eqnarray}
&& \exp(f_\parti(c^{(m)})\vari)q^d \notag \\
&=& \left[ \exp(\phi_m(\ps1\parti, \ps2\parti,\dots, \ps{m}\parti)\vari)
\exp(t_1 \ps1\parti 
   + t_2 \ps2\parti
   + \cdots 
   + t_m \ps{m}\parti) 
\right]_{e^{t_1}=q, t_2=t_3= \cdots=0} \notag \\
&=& \left[ \exp(\phi_m(\Dq, D_2, \dots, D_m)\vari)
\exp(t_1 \ps1\parti
   + t_2 \ps2\parti
   + \cdots 
   + t_m \ps{m}\parti) 
\right]_{e^{t_1}=q, t_2= \cdots= t_m =0}.
\label{eqn:exp q}
\end{eqnarray}
Then by (\ref{shiki:Zhat}), (\ref{eqn:exp q}) and (\ref{eqn:V'}), 
we have
\begin{eqnarray*}
&&\hat{Z}(q,\vari) \\
&=& \left[ \exp(\phi_m(\Dq, D_2,\dots, D_m) \vari) 
\sum_{d\ge0} \sum_{\parti \in \Young} 
\exp(t_1 \ps1\parti
   + t_2 \ps2\parti
   + t_3 \ps3\parti + \cdots)
\right]_{e^{t_1}=q, t_2=t_3= \cdots=0} \\
&=& \left[
\exp(\phi_m(\Dq, D_2,\dots, D_m) \vari) 
V'(q,t_2,t_3,\dots) \right]_{t_2=t_3= \cdots=0} \\
&=& \left[
\exp(\phi_m(\Dq, D_2,\dots, D_m) \vari) 
\exp(\sum_{j=1}^\infty t_j \xi(-j)) V(q,t_2,t_3,\dots) 
\right]_{t_2=t_3= \cdots=0} \\
&=& q^{1/24} \left[
\exp(\phi_m(\Dq + \xi(-1), D_2 + \xi(-2),\dots,D_m+\xi(-m)) \vari) 
V(q,t_2,t_3,\dots) 
\right]_{t_2=t_3= \cdots=0}.
\end{eqnarray*}
Here we have used (\ref{eqn:VV'}).
Then,
\begin{eqnarray}
&& \eta(q) Z(q,\vari) \label{eqn:final}
\\
&=& \eta(q) q^{-1/24} \hat{Z}(q,X) 
\notag \\
&=& \eta(q) \left[
     \exp(\phi_m(\Dq+\xi(-1), D_{2}+\xi(-2),\dots,D_{m}+\xi(-m)) \vari) 
     V(q,t_2,t_3,\dots) \right]_{t_2=t_3= \cdots=0}
\notag \\
&=& \eta(q) 
  \left[
  \exp(\phi_m(\Dq+\xi(-1), D_{2}+\xi(-2),\dots,D_{m}+\xi(-m)) \vari) 
  \sum_{K} A_K(q) \frac{t^K}{K!} \right]_{t_2=t_3=\cdots=0}. \notag
\end{eqnarray}
The coefficient of $\vari^b$ 
on the right-hand side of (\ref{eqn:final})
is equal to the quantity
\[
\frac{1}{b!} \sum_{K}
\eta(q) \left[
\phi_m(\Dq+\xi(-1), D_{2}+\xi(-2),\dots,D_{m}+\xi(-m))^b
A_K(q) \frac{t^K}{K!} \right]_{t_2=t_3=\cdots=0}.
\]
This is a finite sum and belongs to $\QM_*(SL(2,\Z))$,
by Lemma~\ref{lem:eta D}.
Then the right-hand side of (\ref{eqn:final})
is a formal power series in $\vari$
with coefficients in $\QM_*(SL(2,\Z))$.
Hence by (\ref{eqn:eta Z}), we have
\[
\sum_{b\ge 1} F_{1+(m-1)b/2} (q) \vari^b/b!
= \log\left( \eta(q) Z(q,\vari) \right)
= \sum_{l=1}^\infty ( \eta(q) Z(q,\vari) -1)^j (-1)^{j-1} / j.
\]
This shows that
$F_q(q) \in \QM_*(SL(2,\Z))$.
\qed

The special case $m=2$ of our theorem is considered by 
Dijkgraaf \cite{Dijkgraaf}.


Department of Mathematics, Kyushu University,
Hakozaki, Fukuoka, 812-0053, Japan;

ochiai@math.kyushu-u.ac.jp


\begin{thebibliography}{xx}
\bibitem{B}
S. Bloch, 
Zeta values and differential operators on the circle. 
J. Algebra {\bf 182} (1996) 476--500.

\bibitem{BO}
S. Bloch and A. Okounkov,
The character of the infinite wedge representation,
preprint, alg-geom 9712009.

\bibitem{Dijkgraaf}
R. Dijkgraaf, Mirror symmetry and elliptic curves,
Progress in Math. {\bf 129} (1995) 149--163.

\bibitem{Dijkgraaf2}
R. Dijkgraaf,
Chiral deformations of conformal field theories.
Nuclear Phys. B {\bf 493} (1997), 588--612. 

\bibitem{Douglas}
M. R. Douglas, 
Conformal field theory techniques in large $N$ Yang-Mills theory,
Quantum field theory and string theory , 
NATO Adv. Sci. Inst. Ser. B Phys., {\bf 328}, 
(1995)  119--135. 

\bibitem{GH}
P. Griffiths and J. Harris, 
Principles of algebraic geometry. 
Pure and Applied Mathematics. Wiley, New York, 1978.

\bibitem{KZ}
M. Kaneko and D. Zagier,
A generalized Jacobi theta function and quasimodular forms,
Progress in Math. {\bf 129} (1995) 165--172.

\bibitem{Macdonald}
I. G. Macdonald,  Symmetric Functions and Hall Polynomials, 2nd ed.,
Oxford, 1995.

\bibitem{OO}
A. Okounkov and G. Olshanski, 
Shifted Schur functions. II. 
Kirillov's seminar on representation theory,  
Amer. Math. Soc. Transl. Ser. {\bf 2} (1998) 245--271.

\bibitem{Sagan}
B. E. Sagan,  
The Symmetric Group. 
Representations, combinatorial algorithms, and symmetric functions. 
The Wadsworth \& Brooks/Cole, 1991. 

\end{thebibliography}
\end{document}